\documentclass[%
 reprint,
superscriptaddress,
showpacs,
preprintnumbers,
 amsmath,amssymb,
pra,
longbibliography
]{revtex4-1}

\usepackage{graphicx}
\usepackage{dcolumn}
\usepackage{bm}
\usepackage{setspace}	
\usepackage{color}
\usepackage{amsmath}	

\DeclareGraphicsExtensions{.eps}
\graphicspath{{/Users/ogawak/Dropbox/figs/}}

\def\ii{{\mathrm{i}}}
\def\ff{{\mathrm{f}}}
\def\ee{{\mathrm{e}}}
\def\dd{{\mathrm{d}}}

\def\no{{\nonumber}} 

\def\bra#1{\langle #1|}

\def\ket#1{|#1\rangle}

\def\bracket#1{\langle #1 \rangle}
\def\bracketi#1#2{\langle #1 | #2 \rangle}
\def\bracketii#1#2#3{\langle #1 | #2| #3\rangle}

\def\sub#1{_\mathrm{#1}} 

\def\tr{\mathrm{tr}}

\begin{document}


\title{
Operational formulation of weak values without probe systems
}

\author{Kazuhisa Ogawa}
\email{ogawak@ist.hokudai.ac.jp}
\affiliation{%
Graduate School of Information Science and Technology, Hokkaido University, Sapporo 060-0814, Japan
}%
 
\author{Hirokazu Kobayashi}
\affiliation{%
School of System Engineering, Kochi University of Technology, Tosayamada-cho, Kochi 782-8502, Japan
}%

\author{Akihisa Tomita}
\affiliation{%
Graduate School of Information Science and Technology, Hokkaido University, Sapporo 060-0814, Japan
}%

\date{\today}

\begin{abstract}
Weak values are the fundamental values for observables in a pre- and post-selected system.
Weak values are typically measured by weak measurement, in which weak values appear in the change of not the pre- and post-selected system but the probe system. 
This indirect characteristic of weak measurement obscures the meaning of weak values for the pre- and post-selected system, in contrast to conventional physical quantities, which have a clear operational meaning.
In this study, we operationally formulate weak values as the sensitivity of post-selection probability amplitude to small transformation in a pre- and post-selected system.
This formulation of weak values, which is free from the concept of probe shift assumed in weak measurement, gives a direct interpretation of strange weak values for the pre- and post-selected system.
We further explain that this formulation can simplify weak-value measurement experiments because no probe system is required.
\end{abstract}


\maketitle


\begin{spacing}{1.0}

\section{Introduction}

For an observable $\hat{A}$, a quantum system pre- and post-selected in $\ket{\ii}$ and $\ket{\ff}$, respectively, has a complex characteristic value called {\it weak value}:
\begin{align}
 \bracket{\hat{A}}\sub{w}:=\frac{\bracketii{\ff}{\hat{A}}{\ii}}{\bracketi{\ff}{\ii}}.
\end{align}
The weak value was originally introduced as a measurement outcome of {\it weak measurement} for a pre- and post-selected system \cite{PhysRevLett.60.1351}.
In weak measurement, an additional degree of freedom is employed as a probe system besides the pre- and post-selected system to be measured.
A pre-selected system $\ket{\ii}$ weakly interacts with the probe system and is then post-selected in $\ket{\ff}$; after post-selection, the weak value is obtained as a position and momentum shifts of the probe system \cite{jozsa2007complex}.
Weak measurement has been used to study various fundamental problems in quantum mechanics \cite{aharonov1991complete,resch2004experimental,PhysRevLett.102.020404,yokota2009direct,kocsis2011observing,goggin2011violation,denkmayr2014observation,mahler2016experimental} because it provides weak values as intermediate information of the pre- and post-selected systems without disturbing them.
In addition, weak measurement has been applied for precise measurements of magnitudes of weak system--probe interactions
(weak-value amplification) \cite{hosten2008observation,dixon2009ultrasensitive,magana2014amplification,hallaji2017weak}, as well as direct measurements of wavefunctions and pseudo-probability distributions of the system's pre-selected state
\cite{lundeen2011direct,lundeen2012procedure,salvail2013full,kobayashi2014stereographical,malik2014direct,mirhosseini2014compressive,shi2015scan,thekkadath2016direct}.

Weak values can be beyond the range of the observable's eigenvalues, a fact that has been one of the central topics in weak values and weak measurement \cite{PhysRevLett.60.1351,aharonov1991complete,resch2004experimental,PhysRevLett.102.020404,yokota2009direct,hosten2008observation,dixon2009ultrasensitive,magana2014amplification,hallaji2017weak}. 
Such ``strange weak values'' \cite{hosoya2010strange} are intriguing when regarded as an extension of conventional physical quantities, such as classical physical quantities and quantum expectation values.
However, there is a difference between the weak values obtained by weak measurement and the conventional physical quantities in terms of the directness of the measurement procedure.
The conventional physical quantities can be directly measured from the state changes of the measured system itself, and probe systems are not necessarily required.
In contrast, in weak measurement, weak values are observed not in the changes of the pre- and post-selected system but in the shifts of the probe system.
This indirect characteristic of weak measurement makes it difficult to consider weak values as a naive extension of conventional physical quantities and obscures what the weak values represent for the pre- and post-selected system.
Because weak values inherently belong to the pre- and post-selected system, they should be observed in that system's changes.

In this study, we operationally formulate weak values as the sensitivity of the pre- and post-selected system to a small transformation, such as unitary and amplification/attenuation transformations.
The weak value of the derivative of the small transformation is observed as the response of the post-selection probability amplitude change.
This relation also holds for conventional physical quantities; therefore, this formulation allows us to interpret weak values as a natural extension of the conventional physical quantities regardless of the presence of the probe systems.
We apply this formulation to cases of the quantum box problem \cite{aharonov1991complete,resch2004experimental} and the huge weak value of spin \cite{PhysRevLett.60.1351} as examples and examine how the strange weak values can be directly understood as a natural extension of the conventional physical quantities such as probability and spin angular momentum.

In addition, because this formulation does not require extra probe systems, 
the weak-value measurement method according to this formulation is easier to implement than conventional weak measurement.
Thus, we further discuss the applicability of this simple weak-value measurement method and evaluate the performance of this method as a weak-value estimation method in terms of accuracy and precision.

This paper is organized as follows.
In Sec.~\ref{sec:2}, we introduce an operational formulation of weak values and explain how to obtain the real and imaginary parts of weak values experimentally in this formulation.
In Sec.~\ref{sec:3}, we apply this formulation to the cases of the quantum box problem and the huge weak value of spin as examples.
In Sec.~\ref{sec:4}, we examine the performance of this method to show that it can be used to simplify measurement experiments for various applications of weak values.
Finally, we summarize the findings of our study in Sec.~\ref{sec:5}.

\section{Operational formulation of weak values without probe systems}\label{sec:2}

In this section, we operationally formulate the weak values as the sensitivity of the post-selection probability amplitude when a small transformation is set between the pre- and post-selection. 
In Sec.~\ref{sec:2-1}, we first introduce a small linear transformation and provide the general theory of this formulation.
In Secs.~\ref{sec:2-2} and \ref{sec:2-3}, we describe how to obtain real and imaginary parts of weak values in this formulation, respectively.
In Sec.~\ref{sec:2-3.5}, we extend this formulation to the case of the mixed pre- and post-selection and that of the pre-selection only.
In Sec.~\ref{sec:2-4}, we mention the previous studies related to this formulation of weak values.




\subsection{General theory}\label{sec:2-1}

\begin{figure}
\includegraphics[width=8.5cm]{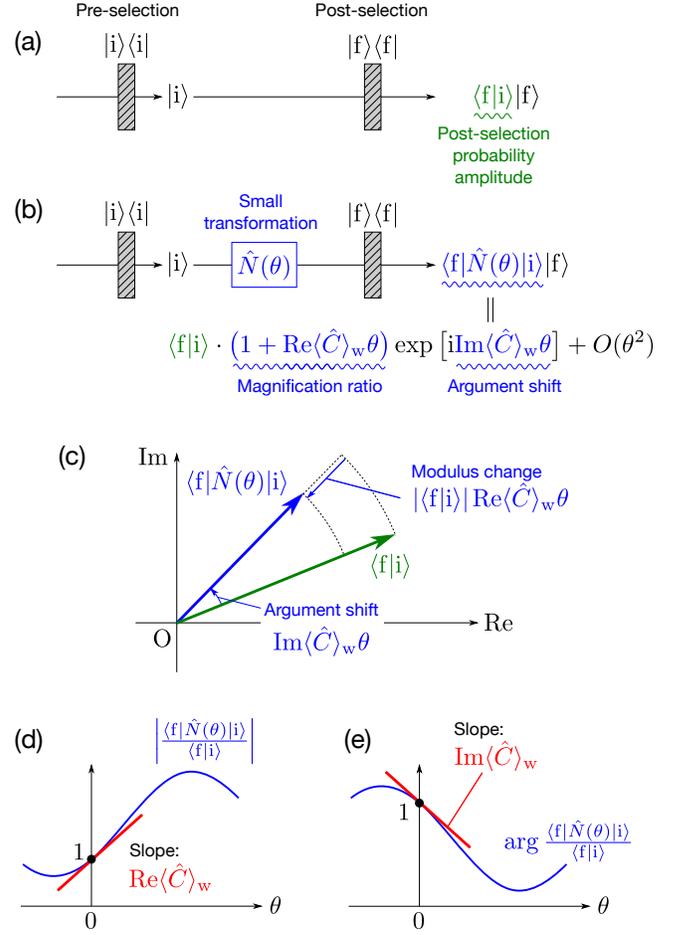}
\caption{
Schematic overview of the proposed operational formulation of weak values.
(a) Pre- and post-selected system $\{\ket{\ii},\ket{\ff}\}$.
The post-selected state has a post-selection probability amplitude $\bracketi{\ff}{\ii}$.
(b) The pre- and post-selected system that includes a small transformation $\hat{N}(\theta)$. 
For its post-selection probability amplitude $\bracketii{\ff}{\hat{N}(\theta)}{\ii}$,
its modulus is magnified $1+\mathrm{Re}\bracket{\hat{C}}\sub{w}\theta$ times, and its argument is shifted by $\mathrm{Im}\bracket{\hat{C}}\sub{w}\theta$ compared with $\bracketi{\ff}{\ii}$.
(c) Relation between $\bracketi{\ff}{\ii}$ and $\bracketii{\ff}{\hat{N}(\theta)}{\ii}$ on the complex plane.
The real and imaginary parts of $\bracket{\hat{C}}\sub{w}$ appear in the differences of the modulus and the argument between them, respectively.
The higher order term $O(\theta^2)$ is not displayed here. 
(d) Relation between $|\bracketii{\ff}{\hat{N}(\theta)}{\ii}/\bracketi{\ff}{\ii}|$ and $\mathrm{Re}\bracket{\hat{C}}\sub{w}$.
$\mathrm{Re}\bracket{\hat{C}}\sub{w}$ is represented as the slope of $|\bracketii{\ff}{\hat{N}(\theta)}{\ii}/\bracketi{\ff}{\ii}|$ at $\theta=0$.
(e) Relation between $\arg(\bracketii{\ff}{\hat{N}(\theta)}{\ii}/\bracketi{\ff}{\ii})$ and $\mathrm{Im}\bracket{\hat{C}}\sub{w}$.
$\mathrm{Im}\bracket{\hat{C}}\sub{w}$ is represented as the slope of $\arg(\bracketii{\ff}{\hat{N}(\theta)}{\ii}/\bracketi{\ff}{\ii})$ at $\theta=0$.
}\label{fig:1-1}
\end{figure}

To derive an operational formulation of the weak value of a normal operator $\hat{C}$ for a pre- and post-selected system, we first introduce the following small linear transformation $\hat{N}(\theta)$.
$\hat{N}(\theta)$ is a normal operator parameterized by a small real parameter $\theta$, which maps a pure state onto another pure state.
We assume that $\hat{N}(\theta)$ satisfies $\hat{N}(0)=\hat{1}$ and can be expanded for $\theta$ as \footnote{We use the notation $O(f(\theta))$ as an operator satisfying $\limsup_{\theta\to 0}||O(f(\theta))/f(\theta)||<\infty$, where $||\cdot||$ is an operator norm.}
\begin{align}
 \hat{N}(\theta)=\hat{1}+\theta\hat{C}+O(\theta^2).\label{eq:17}
\end{align}
$\hat{C}$ is the derivative of $\hat{N}(\theta)$ at $\theta=0$, and we call $\hat{C}$ the {\it generator} of $\hat{N}(\theta)$.
While the exponential form $\exp(\theta\hat{C})$ is an example of $\hat{N}(\theta)$, we do not limit $\hat{N}(\theta)$ to the exponential form here.

We next consider the pre- and post-selected systems $\{\ket{\ii},\ket{\ff}\}$ shown in Figs.~\ref{fig:1-1}(a) and (b).
In the system of Fig.~\ref{fig:1-1}(b), a small linear transformation $\hat{N}(\theta)$ is set between the pre- and post-selection.
After the post-selection, the output states have post-selection probability amplitudes $\bracketi{\ff}{\ii}$ and $\bracketii{\ff}{\hat{N}(\theta)}{\ii}$.
We evaluate the sensitivity of $\bracketii{\ff}{\hat{N}(\theta)}{\ii}$ to $\theta$ by the ratio of $\bracketii{\ff}{\hat{N}(\theta)}{\ii}$ to $\bracketi{\ff}{\ii}$.
This ratio can be expanded for $\theta$ as
\begin{align}
 \frac{\bracketii{\ff}{\hat{N}(\theta)}{\ii}}
{\bracketi{\ff}{\ii}}
&=
1
+\bracket{\hat{C}}\sub{w}\theta
+O(\theta^2).\label{eq:10}
\end{align}
This equation indicates that the derivative of this ratio for $\theta$ at $\theta=0$ corresponds to the weak value of $\hat{C}$:
\begin{align}
\left.
\frac{\dd}{\dd \theta}
\frac{\bracketii{\ff}{\hat{N}(\theta)}{\ii}}
{\bracketi{\ff}{\ii}}
\right|_{\theta=0}
=\bracket{\hat{C}}\sub{w}.\label{eq:26}
\end{align}
In other words, for the pre- and post-selected system $\{\ket{\ii},\ket{\ff}\}$, $\bracket{\hat{C}}\sub{w}$ is formulated as the sensitivity (i.e., rate of change) of the post-selection probability amplitude to the small transformation whose generator is $\hat{C}$.

Moreover, Eqs.~(\ref{eq:10}) and (\ref{eq:26}) can be rewritten as
\footnote{We used the following relations for any complex functions $f(\theta)$ satisfying $f(0)=1$:
\begin{align}
 |f(\theta)|&=\mathrm{Re}f(\theta)+O(\theta^2),\\
\arg f(\theta)&=\mathrm{Im}f(\theta)+O(\theta^2).
\end{align}
}
\begin{align}
 \frac{\bracketii{\ff}{\hat{N}(\theta)}{\ii}}
{\bracketi{\ff}{\ii}}
=\big(1+\mathrm{Re}\bracket{\hat{C}}\sub{w}\theta\big)
&\exp\big[\ii\mathrm{Im}\bracket{\hat{C}}\sub{w}\theta\big]+O(\theta^2),\label{eq:16}\\
\left.
\frac{\dd}{\dd \theta}
\bigg|\frac{\bracketii{\ff}{\hat{N}(\theta)}{\ii}}
{\bracketi{\ff}{\ii}}
\bigg|
\right|_{\theta=0}
&=\mathrm{Re}\bracket{\hat{C}}\sub{w},\label{eq:19}\\
\left.
\frac{\dd}{\dd \theta}
\arg\frac{\bracketii{\ff}{\hat{N}(\theta)}{\ii}}
{\bracketi{\ff}{\ii}}
\right|_{\theta=0}
&=\mathrm{Im}\bracket{\hat{C}}\sub{w}.\label{eq:20}
\end{align}
Equation (\ref{eq:16}) means that the modulus of $\bracketii{\ff}{\hat{N}(\theta)}{\ii}$ is magnified $1+\mathrm{Re}\bracket{\hat{C}}\sub{w}\theta$ times, and its argument is shifted by $\mathrm{Im}\bracket{\hat{C}}\sub{w}\theta$ compared with $\bracketi{\ff}{\ii}$.
This relation is depicted on the complex plane shown in Fig.~\ref{fig:1-1}(c).
Equations (\ref{eq:19}) and (\ref{eq:20}) mean that the real and imaginary parts of $\bracket{\hat{C}}\sub{w}$ are formulated as the sensitivity of the variation of the modulus and argument of the post-selection probability amplitude.
The relations between $|\bracketii{\ff}{\hat{N}(\theta)}{\ii}/\bracketi{\ff}{\ii}|$ and $\mathrm{Re}\bracket{\hat{C}}\sub{w}$ and between $\arg(\bracketii{\ff}{\hat{N}(\theta)}{\ii}/\bracketi{\ff}{\ii})$ and $\mathrm{Im}\bracket{\hat{C}}\sub{w}$ are depicted in Figs.~\ref{fig:1-1}(d) and (e), respectively.

\begin{table}
\caption{
Relations between each type of small transformation and the variation of the modulus and argument of ${\bracketii{\ff}{\hat{N}(\theta)}{\ii}}$.
}
\begin{ruledtabular}\label{tab:1}
\begin{tabular}{ccc}
 &\multicolumn{2}{c}{Variation of ${\bracketii{\ff}{\hat{N}(\theta)}{\ii}}$:}\\
 Small transformation & 
\begin{tabular}{c}
Modulus
\end{tabular}
 & 
\begin{tabular}{c}
Argument 
\end{tabular}
\\ 
\hline
\rule[-15pt]{0pt}{32pt}
\begin{tabular}{c}
Unitary:\\
$\hat{U}(\theta)=\hat{1}+\ii\theta\hat{A}+O(\theta^2)$ 
\end{tabular} 
& $-|\bracketi{\ff}{\ii}|\mathrm{Im}\bracket{\hat{A}}\sub{w}\theta$ 
& $\mathrm{Re}\bracket{\hat{A}}\sub{w}\theta$\\
\rule[-10pt]{0pt}{25pt}
\begin{tabular}{c}
\hspace{-3mm}Amplification/attenuation:\\
$\hat{T}(\theta)=\hat{1}+\theta\hat{B}+O(\theta^2)$  
\end{tabular} 
& $|\bracketi{\ff}{\ii}|\mathrm{Re}\bracket{\hat{B}}\sub{w}\theta$ 
& $\mathrm{Im}\bracket{\hat{B}}\sub{w}\theta$\\
\end{tabular}
\end{ruledtabular}
\end{table}

The remainder of this section notes the special cases of the small transformation $\hat{N}(\theta)$.
The generator $\hat{C}$ is expressed using two Hermite operators $\hat{A}$ and $\hat{B}$ as $\hat{C}=\ii\hat{A}+\hat{B}$.
When $\hat{C}=\ii\hat{A}$, $\hat{N}(\theta)$ becomes a unitary transformation $\hat{U}(\theta):=\hat{1}+\ii\theta\hat{A}+O(\theta^2)$ in the first-order approximation of $\theta$. 
In contrast, when $\hat{C}=\hat{B}$, $\hat{N}(\theta)$ becomes an amplification/attenuation transformation $\hat{T}(\theta):=\hat{1}+\theta\hat{B}+O(\theta^2)$, which amplifies (attenuates) the moduli of the amplitudes in the eigenspaces of $\hat{B}$ whose eigenvalues are positive (negative) when $\theta>0$ (how to realize the amplification/attenuation transformation by unitary processes is explained in Appendix~\ref{sec:appendix1}).
Because of the following relations:
\begin{align}
\mathrm{Re}\bracket{\hat{C}}\sub{w}&=-\mathrm{Im}\bracket{\hat{A}}\sub{w}+\mathrm{Re}\bracket{\hat{B}}\sub{w},\\
\mathrm{Im}\bracket{\hat{C}}\sub{w}&=\mathrm{Re}\bracket{\hat{A}}\sub{w}+\mathrm{Im}\bracket{\hat{B}}\sub{w},
\end{align} 
$\mathrm{Re}\bracket{\hat{A}}\sub{w}$, $\mathrm{Im}\bracket{\hat{A}}\sub{w}$, $\mathrm{Re}\bracket{\hat{B}}\sub{w}$, and $\mathrm{Im}\bracket{\hat{B}}\sub{w}$ appear in the variation of ${\bracketii{\ff}{\hat{N}(\theta)}{\ii}}$ as shown in Table~\ref{tab:1}.

\subsection{How to obtain $\mathrm{Re}\bracket{\hat{C}}\sub{w}$ experimentally
}\label{sec:2-2}

\begin{figure}
\includegraphics[width=8.5cm]{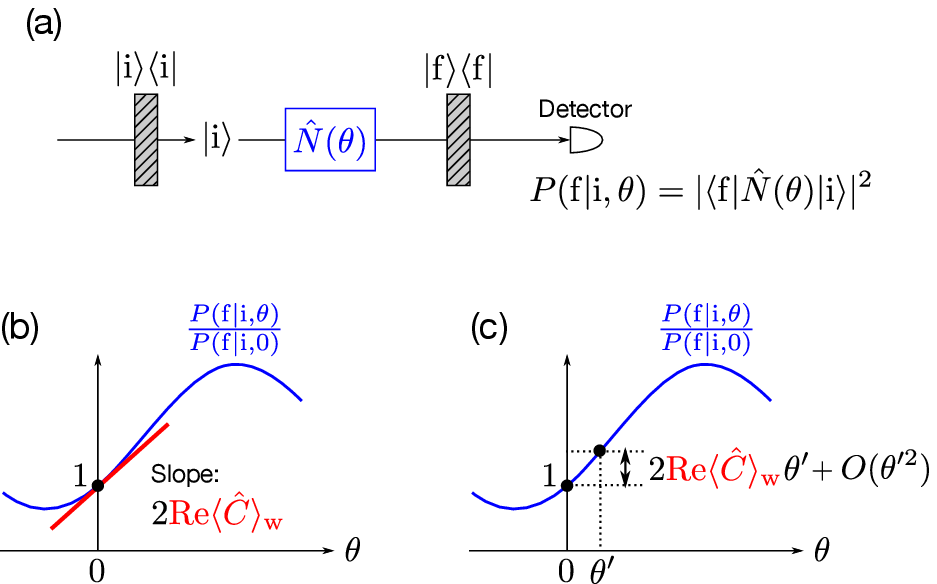}
\caption{
(a) Schematic diagram of the pre- and post-selected system for measuring $\mathrm{Re}\bracket{\hat{C}}\sub{w}$.
(b), (c) Relation between $P(\ff|\ii,\theta)/P(\ff|\ii,0)$ and $\mathrm{Re}\bracket{\hat{C}}\sub{w}$.
(b) $\mathrm{Re}\bracket{\hat{C}}\sub{w}$ is represented as a half of the slope of $P(\ff|\ii,\theta)/P(\ff|\ii,0)$ at $\theta=0$.
(c) Experimentally, $\mathrm{Re}\bracket{\hat{C}}\sub{w}$ is obtained as the difference between 1 and $P(\ff|\ii,\theta')/P(\ff|\ii,0)$ for small $\theta'$.}\label{fig:1-2}
\end{figure}

This section and the next section describe how to obtain $\mathrm{Re}\bracket{\hat{C}}\sub{w}$ and $\mathrm{Im}\bracket{\hat{C}}\sub{w}$ experimentally, respectively. 
According to Eqs.~(\ref{eq:16}) and (\ref{eq:19}), $\mathrm{Re}\bracket{\hat{C}}\sub{w}$ appears in the modulus of $\bracketii{\ff}{\hat{N}(\theta)}{\ii}/\bracketi{\ff}{\ii}$.
The square of this modulus can be obtained by measuring the post-selection probability $P(\ff|\ii,\theta):=|\bracketii{\ff}{\hat{N}(\theta)}{\ii}|^2$ by the system shown in Fig.~\ref{fig:1-2}(a) as 
\begin{align}
\frac{P(\ff|\ii,\theta)}{P(\ff|\ii,0)}
=\frac{|\bracketii{\ff}{\hat{N}(\theta)}{\ii}|^2}{|\bracketii{\ff}{\hat{N}(0)}{\ii}|^2}
=\left|\frac{\bracketii{\ff}{\hat{N}(\theta)}{\ii}}{\bracketi{\ff}{\ii}}\right|^2.
\end{align}
Therefore, $\mathrm{Re}\bracket{\hat{C}}\sub{w}$ can be obtained experimentally by measuring the post-selection probability $P(\ff|\ii,\theta)$ for $\theta=0$ and $\theta\neq0$ as 
\begin{align}
&\frac{P(\ff|\ii,\theta)}{P(\ff|\ii,0)}
=1+2\mathrm{Re}\bracket{\hat{C}}\sub{w}\theta+O(\theta^2),\label{eq:25}\\
&\left.\frac{\dd}{\dd\theta}\frac{P(\ff|\ii,\theta)}{P(\ff|\ii,0)}\right|_{\theta=0}
=2\mathrm{Re}\bracket{\hat{C}}\sub{w}.
 \label{eq:3}
\end{align}
The relation between $P(\ff|\ii,\theta)/P(\ff|\ii,0)$ and $\mathrm{Re}\bracket{\hat{C}}\sub{w}$ is illustrated in Figs.~\ref{fig:1-2}(b) and (c).
Equation~(\ref{eq:3}) indicates that $\mathrm{Re}\bracket{\hat{C}}\sub{w}$ is represented as the slope of $P(\ff|\ii,\theta)/P(\ff|\ii,0)$ at $\theta=0$.
Experimentally, $\mathrm{Re}\bracket{\hat{C}}\sub{w}$ is obtained as the difference between 1 and $P(\ff|\ii,\theta)/P(\ff|\ii,0)$ for small $\theta$.
Remarkably, this method to obtain $\mathrm{Re}\bracket{\hat{C}}\sub{w}$ does not require a probe system, which is unlike conventional weak measurement.
This characteristic can be a practical advantage for weak-value measurement experiments and will be discussed in Sec.~\ref{sec:4}.


  \subsection{How to obtain $\mathrm{Im}\bracket{\hat{C}}\sub{w}$ experimentally
}\label{sec:2-3}

\begin{figure}
\includegraphics[width=8.5cm]{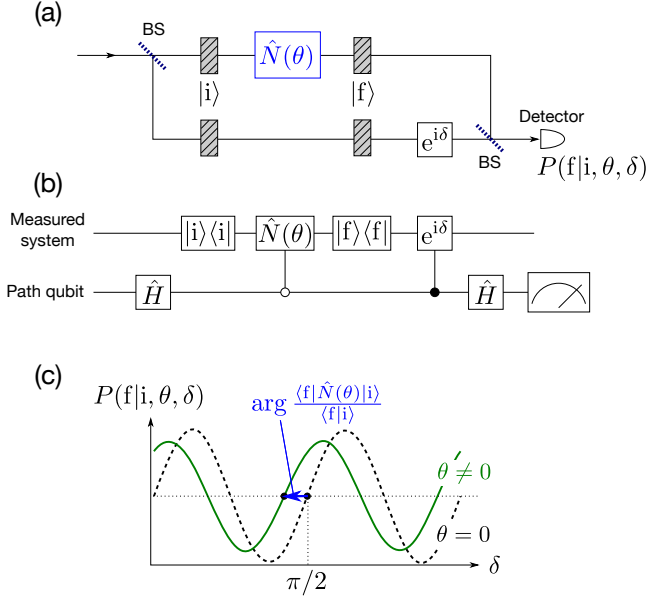}
\caption{
(a) Schematic diagram of the interferometer for measuring $\mathrm{Im}\bracket{\hat{C}}\sub{w}$.
The upper and lower path modes in the interferometer show the added external degrees of freedom.
BS: beam splitter, which works as a Hadamard operator for the external degrees of freedom.  
(b) Quantum circuit representation of the interferometer (a). 
(c) Variation of the detection probability $P(\ff|\ii,\theta,\delta)$ with respect to $\delta$.
The black dashed and green solid lines are the cases for $\theta=0$ and $\theta\neq 0$, respectively.
The phase shift of the interference fringes corresponds to $\arg(\bracketii{\ff}{\hat{N}(\theta)}{\ii}/\bracketi{\ff}{\ii})$, which approximately equals $\mathrm{Im}\bracket{\hat{C}}\sub{w}$.
}\label{fig:1-3}
\end{figure}

Next, we describe how to obtain $\mathrm{Im}\bracket{\hat{C}}\sub{w}$ experimentally in this formulation.
According to Eqs.~(\ref{eq:16}) and (\ref{eq:20}), $\mathrm{Im}\bracket{\hat{C}}\sub{w}$ appears in the argument of $\bracketii{\ff}{\hat{N}(\theta)}{\ii}/\bracketi{\ff}{\ii}$.
This argument can be obtained experimentally by measuring the phase difference between $\arg\bracketii{\ff}{\hat{N}(\theta)}{\ii}$ for $\theta\neq 0$ and $\theta=0$ as
\begin{align}
\arg\bracketii{\ff}{\hat{N}(\theta)}{\ii}-\arg\bracketii{\ff}{\hat{N}(0)}{\ii}
&=\arg\frac{\bracketii{\ff}{\hat{N}(\theta)}{\ii}}{\bracketi{\ff}{\ii}}\no\\
&=\mathrm{Im}\bracket{\hat{C}}\sub{w}\theta+O(\theta^2).\label{eq:24}
\end{align} 
This argument shift, which appears in the global phase of the post-selected state, cannot be measured experimentally with only the system shown in Fig.~\ref{fig:1-2}(a); however, using an interferometer for another degree of freedom as in Figs.~\ref{fig:1-3}(a) and (b), we can measure it experimentally. 
The added external degree of freedom is depicted as the path mode in the interferometer.
The internal degrees of freedom are pre- and post-selected to $\ket{\ii}$ and $\ket{\ff}$, respectively, and in the upper path, the small linear transformation $\hat{N}(\theta)$ is set between the pre- and post-selection. 
The relative phase $\delta$ in the interferometer is varied by the phase shifter $\ee^{\ii\delta}$ in the lower path.  
The post-selection probability $P(\ff|\ii,\theta,\delta)$ is expressed as
\begin{align}
P(\ff|& \ii,\theta,\delta)
=\frac{|\bracketi{\ff}{\ii}|^2}{4}
\Bigg\{
\left|\frac{\bracketii{\ff}{\hat{N}(\theta)}{\ii}}{\bracketi{\ff}{\ii}}\right|^2+1\no\\
&+\left|\frac{\bracketii{\ff}{\hat{N}(\theta)}{\ii}}{\bracketi{\ff}{\ii}}\right|
2\cos\left[\delta-\arg\frac{\bracketii{\ff}{\hat{N}(\theta)}{\ii}}{\bracketi{\ff}{\ii}}\right]
\Bigg\},
\end{align}
which exhibits interference fringes with respect to $\delta$.
For $\theta\neq 0$, the phase of the interference fringe is displaced by $\arg(\bracketii{\ff}{\hat{N}(\theta)}{\ii}/\bracketi{\ff}{\ii})$ from that for $\theta=0$, and the slope of $\arg(\bracketii{\ff}{\hat{N}(\theta)}{\ii}/\bracketi{\ff}{\ii})$ at $\theta=0$ corresponds to $\mathrm{Im}\bracket{\hat{C}}\sub{w}$, as shown in Fig.~\ref{fig:1-3}(c).
Therefore, $\mathrm{Im}\bracket{\hat{C}}\sub{w}$ can be obtained by measuring the interference fringe's phase shift $\arg(\bracketii{\ff}{\hat{N}(\theta)}{\ii}/\bracketi{\ff}{\ii})$ for small $\theta$.
We note that measurement of $\mathrm{Im}\bracket{\hat{C}}\sub{w}$ requires another degree of freedom, which corresponds to a probe system in conventional weak measurement, whereas the formulation of $\mathrm{Im}\bracket{\hat{C}}\sub{w}$ in Eq.~(\ref{eq:20}) does not require it.

\subsection{Case of mixed pre- and post-selection and that of pre-selection only}\label{sec:2-3.5}

Here, we extend the proposed operational formulation of weak values to the case of the mixed pre- and post-selection.
Then, we use this extended formulation to show that the expectation values for a pre-selected system are formulated in the same form as the weak values for a pre- and post-selected system.
This relation indicates that the weak values are natural extensions of the expected values, regardless of the presence of the probe systems.

The weak value of $\hat{C}$ for the system pre- and post-selected in the mixed states $\hat{\rho}\sub{i}$ and $\hat{\rho}\sub{f}$, respectively, is given as 
\begin{align}
\bracket{\hat{C}}\sub{w}=\frac{\tr(\hat{\rho}\sub{f}\hat{C}\hat{\rho}\sub{i})}{\tr(\hat{\rho}\sub{f}\hat{\rho}\sub{i})}.
\end{align}
The formulation proposed in Sec.~\ref{sec:2-1} appears to not be suitable for this case because the post-selection probability amplitude is not defined for mixed states.
However, the measurement methods introduced in Secs.~\ref{sec:2-2} and \ref{sec:2-3} also work to observe the real and imaginary parts of the weak values even for the case of the mixed pre- and post-selection.
In the system shown in Fig.~\ref{fig:1-2}(a), the post-selection probability for the pre- and post-selection $\{\hat{\rho}\sub{i},\hat{\rho}\sub{f}\}$ is given as $P(\ff|\ii,\theta)=\tr\big[\hat{\rho}\sub{f}\hat{N}(\theta)\hat{\rho}\sub{i}\hat{N}^\dag(\theta)\big]$.
Therefore, the following relation same as Eq.~(\ref{eq:3}) holds:
\begin{align}
\left.
 \frac{\dd}{\dd\theta}
\frac{P(\ff|\ii,\theta)}{P(\ff|\ii,0)}
\right|_{\theta=0}
=\left.
 \frac{\dd}{\dd\theta}
\frac{\tr\big[\hat{\rho}\sub{f}\hat{N}(\theta)\hat{\rho}\sub{i}\hat{N}^\dag(\theta)\big]}
{\tr(\hat{\rho}\sub{f}\hat{\rho}\sub{i})}
\right|_{\theta=0}
=2\mathrm{Re}\bracket{\hat{C}}\sub{w}.\label{eq:27}
\end{align}
Similarly, the post-selection probability in the system shown in Fig.~\ref{fig:1-3}(a)
is given as 
\begin{align}
&P(\ff|\ii,\theta,\delta)
=\frac{\tr(\hat{\rho}\sub{f}\hat{\rho}\sub{i})}{4}
\Bigg\{
\frac{\tr\big[\hat{\rho}\sub{f}\hat{N}(\theta)\hat{\rho}\sub{i}\hat{N}^\dag(\theta)\big]}{\tr(\hat{\rho}\sub{f}\hat{\rho}\sub{i})}
+1\no\\
&+
\left|
\frac{\tr\big[\hat{\rho}\sub{f}\hat{N}(\theta)\hat{\rho}\sub{i}\big]}{\tr(\hat{\rho}\sub{f}\hat{\rho}\sub{i})}
\right|
2\cos\left[\delta-
\arg\frac{\tr\big[\hat{\rho}\sub{f}\hat{N}(\theta)\hat{\rho}\sub{i}\big]}{\tr(\hat{\rho}\sub{f}\hat{\rho}\sub{i})}
\right]
\Bigg\}.
\end{align}
The phase difference between the interference fringes when $\theta\neq 0$ and those when $\theta= 0$, $\arg\big\{\tr[\hat{\rho}\sub{f}\hat{N}(\theta)\hat{\rho}\sub{i}]/{\tr(\hat{\rho}\sub{f}\hat{\rho}\sub{i})}\big\}$, shows the following relation same as Eq.~(\ref{eq:20}):
\begin{align}
\left.
\frac{\dd}{\dd\theta}
\arg\frac{\tr\big[\hat{\rho}\sub{f}\hat{N}(\theta)\hat{\rho}\sub{i}\big]}{\tr(\hat{\rho}\sub{f}\hat{\rho}\sub{i})}
\right|_{\theta=0}
=\mathrm{Im}\bracket{\hat{C}}\sub{w}.\label{eq:28}
\end{align}

\begin{figure}
\includegraphics[width=8cm]{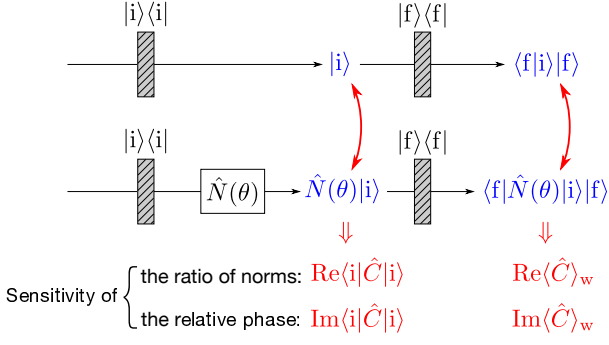}
\caption{Unified interpretation of the expectation values and the weak values in our formulation.
The real (imaginary) part of the expectation values appears in the sensitivity of the ratio of norms (the relative phase) of the states with and without a small transformation in the pre-selected system.
This formulation using the ratio of norms and the relative phase can also be applied to weak values in the pre- and post-selected system.
 }\label{fig:2-D}
\end{figure}

As a special case of the mixed pre- and post-selection, post-selection in the completely mixed state $\hat{1}/d$ ($d$ is the dimension of the system) corresponds to the case of the pre-selection only.
Especially, when the pre-selected state is a pure state $\hat{\rho}\sub{i}=\ket{\ii}\bra{\ii}$, Eq.~(\ref{eq:27}) becomes 
\begin{align}
\left.
 \frac{\dd}{\dd\theta}
\frac{\|\hat{N}(\theta)\ket{\ii}\|}{\|\ket{\ii}\|}
\right|_{\theta=0}
=\mathrm{Re}\bracketii{\ii}{\hat{C}}{\ii}.\label{eq:29}
\end{align}
This equation means that the real part of the expectation value of $\hat{C}$ for the pre-selected state $\ket{\ii}$ appears in the sensitivity to $\theta$ of the ratio of the norms of $\hat{N}(\theta)\ket{\ii}$ for $\theta\neq 0$ and $\theta=0$.
The formulation of $\mathrm{Re}\bracket{\hat{C}}\sub{w}$ in Eq.~(\ref{eq:19}), which looks like a different form from that of Eq.~(\ref{eq:29}), can also be rewritten as the ratio of the norms after the post-selection as 
\begin{align}
\left.
 \frac{\dd}{\dd\theta}
\frac{\|\bracketii{\ff}{\hat{N}(\theta)}{\ii}\ket{\ff}\|}
{\|\bracketi{\ff}{\ii}\ket{\ff}\|}
\right|_{\theta=0}
=\mathrm{Re}\bracket{\hat{C}}\sub{w}.
\end{align}
Besides, Eq.~(\ref{eq:28}) for the pre-selected system $\ket{\ii}$ becomes
\begin{align}
\left.
 \frac{\dd}{\dd\theta}
\arg\bra{\ii}\big[\hat{N}(\theta)\ket{\ii}\big]
\right|_{\theta=0}
=\mathrm{Im}\bracketii{\ii}{\hat{C}}{\ii}.\label{eq:30}
\end{align}
This equation indicates that the imaginary part of the expectation value $\bracket{\hat{C}}$ appears in the sensitivity to $\theta$ of the relative phase of $\hat{N}(\theta)\ket{\ii}$ to $\ket{\ii}$, where the relative phase of $\ket{\psi}$ to $\ket{\phi}$ is defined as $\arg\bracketi{\phi}{\psi}$ \cite{pancharatnam1956generalized}.
The formulation of $\mathrm{Im}\bracket{\hat{C}}\sub{w}$ in Eq.~(\ref{eq:20}) can also be rewritten as the sensitivity of the relative phase after the post-selection as 
\begin{align}
\left.
 \frac{\dd}{\dd\theta}
\arg\big[\bra{\ff}\bracketi{\ii}{\ff}\big]
\big[\bracketii{\ff}{\hat{N}(\theta)}{\ii}\ket{\ff}\big]
\right|_{\theta=0}
=\mathrm{Im}\bracket{\hat{C}}\sub{w}.
\end{align}
In this manner, a unified interpretation of the expectation values and the weak values is given in our formulation, as shown in Fig.~\ref{fig:2-D}, and the weak values are interpreted as natural extensions of the expected values in our formulation, which requires no probe systems.

\subsection{Related previous studies and novelty of our formulation}\label{sec:2-4}

The operational formulation of weak values as the sensitivity of the system to a small transformation has been mentioned partially in previous studies \cite{hofmann2011uncertainty,denkmayr2014observation,yokota2019real,ho2016interpretation,RevModPhys.86.307,qiu2017precision,PhysRevA.93.042124,tamate2009geometrical}.
Our formulation described so far is a generalization of them.
Here, we refer to each of them and explain the novelty of our formulation.

The relation between the modulus change of post-selection probability amplitude and the weak value of the generator of the small transformation in Eq.~(\ref{eq:19}) (or Eq.~(\ref{eq:3})) has been mentioned in Refs.~\cite{hofmann2011uncertainty,denkmayr2014observation,yokota2019real,ho2016interpretation,RevModPhys.86.307,qiu2017precision,PhysRevA.93.042124}.
Refs.~\cite{hofmann2011uncertainty,yokota2019real,ho2016interpretation,RevModPhys.86.307,qiu2017precision} assume that the small transformation is the unitary transformation in the form of $\exp(\ii\theta\hat{A})$ and show that $\mathrm{Im}\bracket{\hat{A}}\sub{w}$ appears in the modulus change of post-selection probability amplitude.
Refs.~\cite{denkmayr2014observation,yokota2019real} assume the attenuation transformation in the form of $\exp(\theta\hat{B})$ and show that $\mathrm{Re}\bracket{\hat{B}}\sub{w}$ appears there.
In contrast, the relation between the argument shift of post-selection probability amplitude and the weak value of the generator of the small transformation in Eq.~(\ref{eq:20}) has been mentioned in Refs.~\cite{tamate2009geometrical,ho2016interpretation,PhysRevA.93.042124}.
These studies assume that the small transformation is the unitary transformation in the form of $\exp(\ii\theta\hat{A})$ and show that $\mathrm{Re}\bracket{\hat{A}}\sub{w}$ appears in the argument shift of post-selection probability amplitude.
Also, we note that Refs.~\cite{hofmann2011uncertainty,denkmayr2014observation,yokota2019real,tamate2009geometrical,qiu2017precision} do not assume the use of probe systems like our formulation, whereas Refs.~\cite{ho2016interpretation,RevModPhys.86.307,PhysRevA.93.042124} assume the von Neumann interaction with probe systems.

Compared with the related previous studies, our formulation of weak values has the following novelty:
(i) We consider a small transformation to be not only the unitary transformation $\exp(\ii\theta\hat{A})$ and the attenuation transformation $\exp(\theta\hat{B})$ but also the general form of Eq.~(\ref{eq:17}).
It includes, for example, a mixture of the unitary and attenuation transformations, an amplification transformation, and a linearly parameterized attenuation transformation $\hat{1}-\theta\hat{C}$. 
(ii) We point out the fact that when a small amplification/attenuation transformation $\hat{1}+\theta\hat{B}+O(\theta^2)$ is set between pre- and post-selection, the imaginary part of the weak value of $\hat{B}$ appears in the argument shift of the post-selection probability amplitude, as shown in the bottom right of the table~\ref{tab:1}.
(iii) Our formulation is applicable to the systems pre- and post-selected in mixed states, which include pre-selected-only systems.
Consequently, the weak values are interpreted as natural extensions of the expected values even without probe systems.

\section{Examples of direct interpretation of strange weak values by the operational formulation}\label{sec:3}

In this section, we apply our formulation of weak values to the examples of the strange weak values.
We deal with the case of the quantum box problem \cite{aharonov1991complete,resch2004experimental} in Sec.~\ref{sec:3-2}, and the case of the huge weak value of the spin of a spin-1/2 particle \cite{PhysRevLett.60.1351} in Sec.~\ref{sec:3-1}.
In both sections, we explain how the conventional physical quantities---classical physical quantities and/or quantum expectation values---and the weak values are interpreted directly by our formulation.






 \subsection{Case of quantum box problem}\label{sec:3-2}

\begin{figure}[t]
\begin{center}
\includegraphics[width=8.5cm]{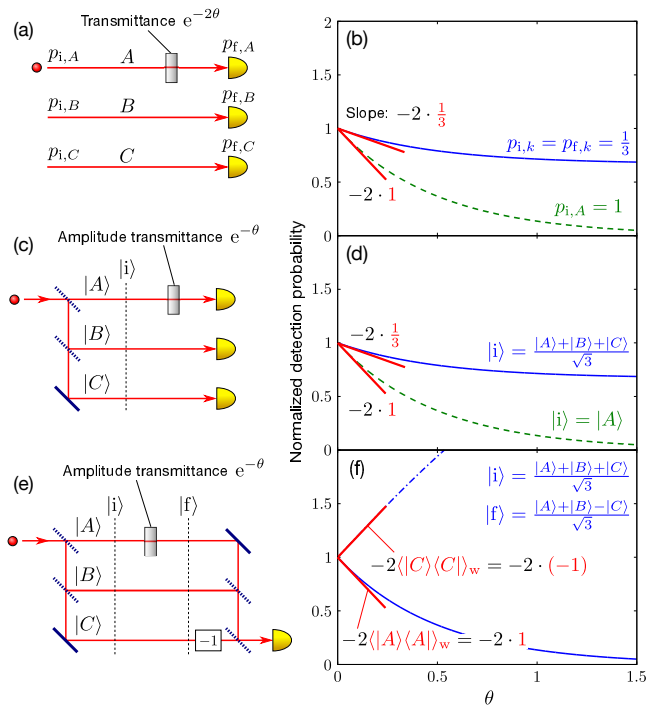}
\caption{
(a), (b) Pre- and post-selected classical system and its normalized detection probability when a probabilistic shutter is set on the path $A$.
The solid blue curve shows that for the case of $p_{\ii,k}=p_{\ff,k}=1/3$ for all $k$.
Its slope at $\theta=0$, $-2\cdot(1/3)$, indicates that the conditional existence probability on the path $A$ of the post-selected particle is $1/3$.  
The green dashed curve shows that for the case of $p_{\ii,A}=1$ and its slope at $\theta=0$, $-2\cdot 1$, indicates that the particle has existed on the path $A$ with a probability of $1$.
(c), (d) Quantum system pre-selected in $\ket{\ii}$ and its detection probability when an attenuator is set on the path $A$.
The solid blue curve and green dashed curve show those for the case of $\ket{\ii}=(\ket{A}+\ket{B}+\ket{C})/\sqrt{3}$ and $\ket{\ii}=\ket{A}$, respectively.
As in graph (b), their slopes at $\theta=0$ indicate the conditional existence probability on the path $A$.  
(e) Quantum system pre- and post-selected in $\ket{\ii}=(\ket{A}+\ket{B}+\ket{C})/\sqrt{3}$ and $\ket{\ff}=(\ket{A}+\ket{B}-\ket{C})/\sqrt{3}$, respectively.
(f) Normalized detection probabilities when an attenuator is set on the path $A$ (solid blue curve) and $C$ (dash-dot blue curve).
The slope of the solid blue curve at $\theta=0$ is $-2\cdot 1$, which looks as if the particle has existed on the path $A$ with a probability of $1$.
The slope of the dash-dot blue curve at $\theta=0$ is $-2\cdot(-1)$, which looks as if a ``negative'' particle has existed on the path $C$.
}\label{fig:3}
\end{center}
\end{figure}

Here, we analyze the quantum box problem using a small attenuation transformation, as an example of the case where the real part of the weak value appears in the sensitivity of the post-selection probability mentioned in Sec.~\ref{sec:2-2}.
The quantum box problem was proposed as a gedankenexperiment where a pre- and post-selected system shows a curious result \cite{aharonov1991complete}, and later, its three box version (three box problem) was experimentally demonstrated using optical weak measurement \cite{resch2004experimental}.
In the three box problem, the particle can be a superposition of three orthogonal path states $\ket{A}$, $\ket{B}$, and $\ket{C}$.
When the particle is pre- and post-selected in $\ket{\ii}=(\ket{A}+\ket{B}+\ket{C})/\sqrt{3}$ and $\ket{\ff}=(\ket{A}+\ket{B}-\ket{C})/\sqrt{3}$, respectively, the weak values of the projection operators of the paths, which are called {\it weak probabilities}, become 
\begin{align}
\bracket{\ket{A}\bra{A}}\sub{w}=\bracket{\ket{B}\bra{B}}\sub{w}=1,\ \text{and}\ 
\bracket{\ket{C}\bra{C}}\sub{w}=-1.
\end{align}
Unlike ordinary probabilities, the weak probabilities can be arbitrary complex values out of $[0,1]$ while satisfying that the sum of them is one.
Here, we explain that in the proposed formulation, such strange weak probabilities can also be understood in a similar way to the existence probabilities of particles in classical and pre-selected quantum systems.

First, let us consider the classical pre- and post-selected system shown in \ref{fig:3}(a), in which the particle is prepared in the path $j$ with a probability of $p_{\ii,j}$ $(\sum_jp_{\ii,j}=1)$ at pre-selection and is detected in path $j$ with a probability of $p_{\ff,j}$ at post-selection.
The detection probability $P(\ff|\ii)$ is given as
\begin{align}
P(\ff|\ii)=p_{\ii,A}p_{\ff,A}+p_{\ii,B}p_{\ff,B}+p_{\ii,C}p_{\ff,C}.
\end{align}
Next, we put a probabilistic shutter which passes the particle with a probability of $\ee^{-2\theta}$ ($\theta$ is a non-negative real parameter) on the path $k$.
In this case, the detection probability $P(\ff|\ii,k,\theta)$ is given as
\begin{align}
P(\ff|\ii,k,\theta)
&=P(\ff|\ii)-(1-\ee^{-2\theta})p_{\ii,k}p_{\ff,k}.
\end{align}
The normalized detection probability $P(\ff|\ii,k,\theta)/P(\ff|\ii)$ has the following derivative with respect to $\theta$ at $\theta=0$:
\begin{align}
 \left.\frac{\dd}{\dd\theta}\frac{P(\ff|\ii,k,\theta)}{P(\ff|\ii)}\right|_{\theta=0}=-2\frac{p_{\ii,k}p_{\ff,k}}{P(\ff|\ii)}.\label{eq:22}
\end{align}
The right-hand side (${p_{\ii,k}p_{\ff,k}}/{P(\ff|\ii)}$) means the conditional existence probability on the path $k$ of the post-selected particles, and it appears in the sensitivity of the normalized detection probability (${P(\ff|\ii,k,\theta)}/{P(\ff|\ii)}$) with respect to $\theta$ at $\theta=0$.
The relation between ${P(\ff|\ii,A,\theta)}/{P(\ff|\ii)}$ and ${p_{\ii,A}p_{\ff,A}}/{P(\ff|\ii)}$ is depicted in Fig.~\ref{fig:3}(b).
For example, when $p_{\ii,k}=p_{\ff,k}=1/3$ for all $k$, the particle's conditional existence probability is $1/3$, and the variation of ${P(\ff|\ii,A,\theta)}/{P(\ff|\ii)}$ (blue solid curve) has a slope of $-2\cdot(1/3)$ at $\theta=0$.
In another example, when $p_{\ii,A}=1$, the particle has existed on the path $A$ with a probability of $1$, and the variation of ${P(\ff|\ii,A,\theta)}/{P(\ff|\ii)}$ (green dashed curve) has a slope of $-2\cdot1$ at $\theta=0$.
These results can be intuitively understood: a path with a higher existence probability should be more affected by the shutter, and the reduction rate of the normalized detection probability should be larger.

This intuitive relation also holds for pre-selected quantum systems. 
Next, we consider the quantum system pre-selected in $\ket{\ii}$ shown in Fig.~\ref{fig:3}(c).
Because post-selection is not performed, the detection probability is $P(\ff|\ii)=\|\ket{\ii}\|^2=1$.
The filter on the path $k$ with amplitude transmittance $\ee^{-\theta}$, which corresponds to the probabilistic shutter in the classical case, is represented as an attenuation transformation $\exp(-\theta\ket{k}\bra{k})$.
The detection probability $P(\ff|\ii,k,\theta)$ is given as
\begin{align}
P(\ff|\ii,k,\theta)
&=\|\exp(-\theta\ket{k}\bra{k})\ket{\ii}\|^2\no\\
&=1-(1-\ee^{-2\theta})\bracketi{\ii}{k}\bracketi{k}{\ii},
\end{align}
and the derivative of the normalized detection probability $P(\ff|\ii,k,\theta)/P(\ff|\ii)$ with respect to $\theta$ at $\theta=0$ is given as
\begin{align}
 \left.\frac{\dd}{\dd\theta}\frac{P(\ff|\ii,k,\theta)}{P(\ff|\ii)}\right|_{\theta=0}
&=\left.\frac{\dd}{\dd\theta}
\frac{\|\exp(-\theta\ket{k}\bra{k})\ket{\ii}\|^2}{\|\ket{\ii}\|^2}
\right|_{\theta=0}\no\\
&=-2\bracketi{\ii}{k}\bracketi{k}{\ii},\label{eq:23}
\end{align}
Therefore, the existence probability of the particle on the path $k$ ($\bracketi{\ii}{k}\bracketi{k}{\ii}$) appears in the sensitivity of the normalized detection probability (${P(\ff|\ii,k,\theta)/P(\ff|\ii)}$) with respect to $\theta$ at $\theta=0$, as mentioned in Sec.~\ref{sec:2-3.5}.
The relation between ${P(\ff|\ii,A,\theta)}/{P(\ff|\ii)}$ and $\bracketi{\ii}{A}\bracketi{A}{\ii}$ is depicted in Fig.~\ref{fig:3}(d).
The variation of $P(\ff|\ii,A,\theta)/P(\ff|\ii)$ when $\ket{\ii}=(\ket{A}+\ket{B}+\ket{C})/\sqrt{3}$ (solid blue curve) and $\ket{\ii}=\ket{A}$ (green dashed curve) show the same curves as those of the classical system in Fig.~\ref{fig:3}(b).

Finally, we consider the quantum system pre- and post-selected in $\ket{\ii}$ and $\ket{\ff}$, respectively, shown in Fig.~\ref{fig:3}(e).
The detection probability with the attenuation filter on the path $k$ is represented as 
\begin{align}
P(\ff|\ii,k,\theta)
&=|\bracketii{\ff}{\exp(-\theta\ket{k}\bra{k})}{\ii}|^2\no\\
&=|\bracketi{\ff}{\ii}|^2\big[1-(1-\ee^{-\theta})2\mathrm{Re}\bracket{\ket{k}\bra{k}}\sub{w}\no\\
&\hspace{1.5cm} +(1-\ee^{-\theta})^2|\bracket{\ket{k}\bra{k}}\sub{w}|^2\big].
\end{align}
The derivative of the normalized detection probability $P(\ff|\ii,k,\theta)/P(\ff|\ii)$ with respect to $\theta$ at $\theta=0$ is given as
\begin{align}
 \left.\frac{\dd}{\dd\theta}\frac{P(\ff|\ii,k,\theta)}{P(\ff|\ii)}\right|_{\theta=0}=-2\mathrm{Re}\bracket{\ket{k}\bra{k}}\sub{w},\label{eq:21}
\end{align}
which has the same form as Eq.~(\ref{eq:3}) in which $\hat{C}=-\ket{k}\bra{k}$.
The weak probability on the path $k$ ($\bracket{\ket{k}\bra{k}}\sub{w}$) appears in the rate of change of the normalized detection probability (${P(\ff|\ii,k,\theta)}/{P(\ff|\ii)}$) with respect to $\theta$ at $\theta=0$.
By identifying Eq.~(\ref{eq:21}) with Eqs.~(\ref{eq:22}) and (\ref{eq:23}), the weak probability can be understood in the same manner as the ordinary existence probability of the particle.
When $\ket{\ii}=(\ket{A}+\ket{B}+\ket{C})/\sqrt{3}$ and $\ket{\ff}=(\ket{A}+\ket{B}-\ket{C})/\sqrt{3}$, the variation of ${P(\ff|\ii,k,\theta)}/{P(\ff|\ii)}$ for $k=A$ and $C$ are illustrated as the solid and dash-dot blue curves, respectively, in Fig.~\ref{fig:3}(f).
The solid blue curve, whose slope at $\theta=0$ is $-2\cdot 1$, has the same shape as the green dashed lines in Figs.~\ref{fig:3}(b) and (d); therefore, the variation of ${P(\ff|\ii,A,\theta)}/{P(\ff|\ii)}$ looks as if the particle exists on path $A$ with a probability of $1$.
The same argument holds for the path $B$.
In contrast, the dash-dot blue curve, whose slope at $\theta=0$ is $-2\cdot(-1)$, has no counterparts in the classical and pre-selected quantum systems.
The normalized detection probability ${P(\ff|\ii,C,\theta)}/{P(\ff|\ii)}$ is amplified despite the attenuation transformation $\exp(-\theta\ket{C}\bra{C})$, and it looks as if a ``negative'' particle exists on the path $C$ \cite{yokota2019real}.
In this manner, the strange weak probabilities can be understood in a similar way as the correspondence between the particle's conditional existence probability and the response of the detection probability to an attenuation transformation in classical or pre-selected quantum systems.

 \subsection{Case of the large weak value of the spin of a spin-1/2 particle}
\label{sec:3-1}

\begin{figure*}
\includegraphics[width=18cm]{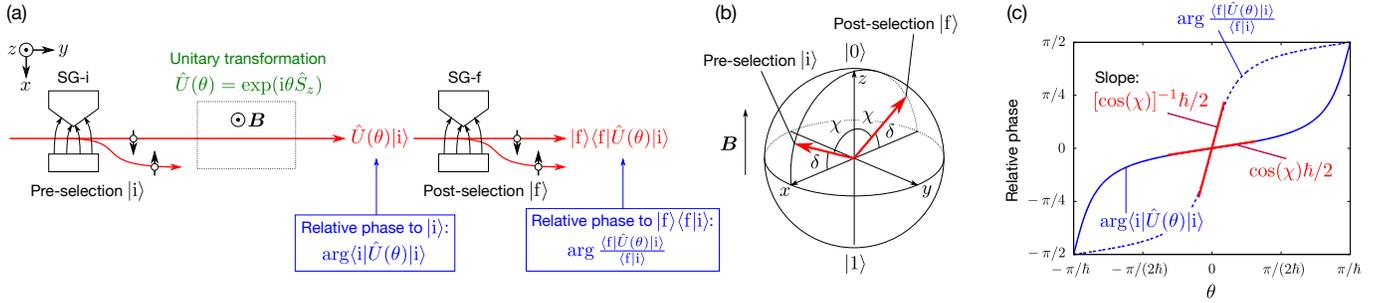}
\caption{
(a) Schematic setup considered in Sec.~\ref{sec:3-1}.
The spin-1/2 particles are pre- and post-selected in $\ket{\ii}$ and $\ket{\ff}$ by the Stern-Gerlach devices SG-i and SG-f, respectively.
Between the pre- and post-selection, the particles undergo a unitary transformation $\hat{U}(\theta)=\exp(\ii\theta\hat{S}_z)$ by a uniform magnetic field in $z$-direction.
$\theta$ corresponds to the strength of the unitary transformation, which includes the interaction time $t$ and the strength of the magnetic field $B$.
The particles' states before the post-selection, $\hat{U}(\theta)\ket{\ii}$ and $\ket{\ii}$, show the relative phase $\arg\bracketii{\ii}{\hat{U}(\theta)}{\ii}$, and those after the post-selection, $\ket{\ff}\bra{\ff}\hat{U}(\theta)\ket{\ii}$ and $\ket{\ff}\bracketi{\ff}{\ii}$, show the relative phase $\arg(\bracketii{\ff}{\hat{U}(\theta)}{\ii}/\bracketi{\ff}{\ii})$.
(b) Bloch sphere representation of the pre- and post-selected states.
When $\delta$ is small, the pre- and post-selected states are nearly orthogonal.
(c) Variation of the relative phases $\arg\bracketii{\ii}{\hat{U}(\theta)}{\ii}$ (blue solid line) and $\arg(\bracketii{\ff}{\hat{U}(\theta)}{\ii}/\bracketi{\ff}{\ii})$ (blue dashed line) when $\chi=7\pi/16$ ($\delta=\pi/16$).
The slope of $\arg\bracketii{\ii}{\hat{U}(\theta)}{\ii}$ at $\theta=0$ corresponds to the expectation value of the spin angular momentum for $\ket{\ii}$: $\cos(\chi)\hbar/2\in[-\hbar/2,\hbar/2]$.
The slope of $\arg(\bracketii{\ff}{\hat{U}(\theta)}{\ii}/\bracketi{\ff}{\ii})$ at $\theta=0$ is $[\cos(\chi)]^{-1}\hbar/2\approx\delta^{-1}\hbar/2$, which becomes arbitrarily large when $\delta$ is small.
When this slope is identified with their spin angular momentum, it seems like the post-selected particles have a huge spin angular momentum. 
}\label{fig:2}
\end{figure*}

In the seminal paper of weak values \cite{PhysRevLett.60.1351}, Aharonov {\it et al.}~proposed that the weak value of the spin-$z$ of a spin-1/2 particle can be $100\hbar$ under appropriate pre- and post-selection.
We next analyze this huge spin weak value using a small unitary transformation, as an example of the case where the real part of the weak value appears in the sensitivity of the argument of the post-selection probability amplitude mentioned in Sec.~\ref{sec:2-3}.
Based on our formulation of weak values, we explain that this huge spin weak value can also be observed in the change of the spin system itself under a magnetic field, in a method similar to the expectation value of the spin angular momentum in a pre-selected-only spin system.
Note that their classical counterpart is not considered here because the argument cannot be defined in the classical system.

First, we consider that a spin-1/2 system pre-selected in $\ket{\ii}$ interacts slightly with the $z$-directional uniform constant magnetic field $B$ as shown in fig.~\ref{fig:2}(a).
$\hat{S}_z$ is the $z$-directional spin operator of a spin-1/2 particle, and $\ket{0}$ and $\ket{1}$ are the eigenstates of $\hat{S}_z$ for eigenvalues $+\hbar/2$ and $-\hbar/2$, respectively.
The interaction Hamiltonian is $\hat{H}=-\gamma \hat{S}_z B$, where $\gamma$ is the gyromagnetic ratio, and the time evolution unitary operator for time $t$ is $\exp(\ii\gamma \hat{S}_z Bt/\hbar)=\exp(\ii\theta \hat{S}_z)=:\hat{U}(\theta)$, where $\theta:=\gamma B t/\hbar$.
After the interaction, the spin state is transformed into $\hat{U}(\theta)\ket{\ii}$.
While the unitary transformation $\hat{U}(\theta)$ does not change the norm of $\hat{U}(\theta)\ket{\ii}$, the global phase of $\hat{U}(\theta)\ket{\ii}$ changes depending on the spin $z$-component of $\ket{\ii}$.
The relative phase of $\hat{U}(\theta)\ket{\ii}$ to $\ket{\ii}$ shows the following relations:
\begin{align}
\left.\frac{\dd}{\dd\theta}
{\arg\bra{\ii}\big[\hat{U}(\theta)\ket{\ii}\big]}
\right|_{\theta=0}
=\mathrm{Im}\bracketii{\ii}{(\ii\hat{S}_z)}{\ii}
=\bracketii{\ii}{\hat{S}_z}{\ii}.\label{eq:12}
\end{align}
Therefore, the expectation value of $\hat{S}_z$ for the pre-selected state $\ket{\ii}$ ($\bracketii{\ii}{\hat{S}_z}{\ii}$) appears in the sensitivity to $\theta$ of the relative phase of $\hat{U}(\theta)\ket{\ii}$ to $\ket{\ii}$ ($\arg\bracketii{\ii}{\hat{U}(\theta)}{\ii}$) under an external effect (the magnetic field), as mentioned in Sec.~\ref{sec:2-3.5}.
The variation of $\arg\bracketii{\ii}{\hat{U}(\theta)}{\ii}$ for $\ket{\ii}=\cos(\chi/2)\ket{0}+\sin(\chi/2)\ket{1}\ (\chi\in[0,\pi])$ is shown in fig.~\ref{fig:2}(c).
Its slope at $\theta=0$ is $\bracketii{\ii}{\hat{S}_z}{\ii}=\cos(\chi)\hbar/2$, which is bounded in the range of the eigenvalues of $\hat{S}_z$, $[-\hbar/2,\hbar/2]$.

Next, we consider a spin-1/2 system pre- and post-selected in $\ket{\ii}$ and $\ket{\ff}$, respectively, as shown in fig.~\ref{fig:2}(a).
The post-selected (unnormalized) state becomes $\ket{\ff}\bracketii{\ff}{\hat{U}(\theta)}{\ii}$, and its relative phase to $\ket{\ff}\bracketi{\ff}{\ii}$ shows the following relations:
\begin{align}
\left.\frac{\dd}{\dd\theta}
\arg\big[{\bracketi{\ii}{\ff}\bra{\ff}}\big]\big[\ket{\ff}\bracketii{\ff}{\hat{U}(\theta)}{\ii}\big]
\right|_{\theta=0}
&=\left.\frac{\dd}{\dd\theta}\arg\frac{\bracketii{\ff}{\hat{U}(\theta)}{\ii}}{\bracketi{\ff}{\ii}}\right|_{\theta=0}\no\\
&=\mathrm{Re}\bracket{\hat{S}_z}\sub{w}.\label{eq:13}
\end{align}
Equation~(\ref{eq:13}) has the same form as that of Eq.~(\ref{eq:20}) in which $\hat{C}=\ii\hat{S}_z$ and indicates that $\mathrm{Re}\bracket{\hat{S}_z}\sub{w}$ corresponds to the sensitivity to $\theta$ of the relative phase of $\ket{\ff}\bracketii{\ff}{\hat{U}(\theta)}{\ii}$ to $\ket{\ff}\bracketi{\ff}{\ii}$ \cite{tamate2009geometrical}.
Unlike $\bracketii{\ii}{\hat{S}\sub{z}}{\ii}$ in Eq.~(\ref{eq:12}), $\mathrm{Re}\bracket{\hat{S}_z}\sub{w}$ can be an arbitrary real number. 
For example, if $\ket{\ii}=\cos(\chi/2)\ket{0}+\sin(\chi/2)\ket{1}$ and $\ket{\ff}=\cos(\chi/2)\ket{0}-\sin(\chi/2)\ket{1}$ $(\chi\in[0,\pi])$ as shown in Fig.~\ref{fig:2}(b), $\mathrm{Re}\bracket{\hat{S}_z}\sub{w}=[\cos(\chi)]^{-1}\hbar/2$.
When $\chi=\pi/2-\delta$ and $0<\delta\ll 1$, $\mathrm{Re}\bracket{\hat{S}_z}\sub{w}\approx \delta^{-1}\hbar/2$, which is much larger than $\hbar/2$.
In this case, the relative phase $\arg(\bracketii{\ff}{\hat{U}(\theta)}{\ii}/\bracketi{\ff}{\ii})$ changes the sensitively to the change of $\theta$ as shown in Fig.~\ref{fig:2}(c).
By identifying Eq.~(\ref{eq:13}) with Eq.~(\ref{eq:12}), the particles look as if they have a huge angular momentum $\delta^{-1}\hbar/2$.
In this manner, the huge weak value can be interpreted as a natural extension of the expectation value of the spin angular momentum.
We note that the measurement of these relative phase requires an additional degree of freedom as mentioned in Sec.~\ref{sec:2-3}.

\section{Performance evaluation of weak-value measurement method without using probe systems}\label{sec:4}

In the previous sections, we illustrated the conceptual advantage of the proposed operational formulation of weak values so that the strange weak values can be understood directly.
Here, we also note that the measurement method of weak values according to this formulation has a practical advantage that this measurement systems can be simplified compared with the original weak measurement because extra probe systems are not required.
To clarify the applicability of our measurement method to various weak-value measurement experiments, we discuss the following three points.
In Sec.~\ref{sec:4-1}, we describe how to apply our measurement method to weak-value amplification.
In Sec.~\ref{sec:4-2}, we evaluate the performance of our measurement method as a weak-value estimation method in terms of accuracy and precision.
In Sec.~\ref{sec:4-3}, we mention the weak-value measurement methods other than weak measurement previously reported and compare the advantages of these methods and our measurement method.


 \subsection{Weak-value amplification in the proposed formulation}\label{sec:4-1}

\begin{figure}[t]
\begin{center}
\includegraphics[width=8.5cm]{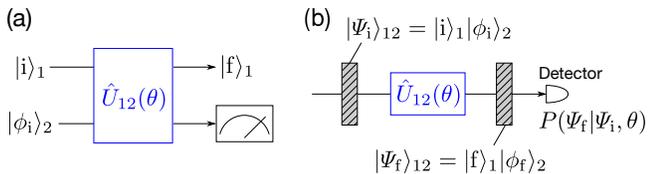}
\caption{(a) Quantum system for weak-value amplification.
The upper system (system 1) is pre- and post-selected in $\ket{\ii}_1$ and $\ket{\ff}_1$, respectively.
The lower system (system 2) is a probe system pre-selected in $\ket{\phi\sub{i}}_2$, and after the interaction $\hat{U}_{12}(\theta)$, the probe shift is measured.
When $|_1\bracketi{\ff}{\ii}_1|\ll 1$, $|\bracket{\hat{A}_1}\sub{w}|$ becomes large; therefore, the probe shift can be detected even if $\theta$ is very small.
(b) Quantum system in our measurement method.
Systems 1 and 2 are considered as a composite system pre- and post-selected in $\ket{\varPsi\sub{i}}_{12}$ and $\ket{\varPsi\sub{f}}_{12}$, respectively.
$\theta$ is estimated from the variation of the post-selection probability $P(\varPsi\sub{f}|\varPsi\sub{i},\theta)$, which can be detected even if $\theta$ is very small when $|_{12}\bracketi{\varPsi\sub{f}}{\varPsi\sub{i}}_{12}|\ll 1$.
}\label{fig:4-1}
\end{center}
\end{figure}

Weak-value amplification \cite{hosten2008observation,dixon2009ultrasensitive,magana2014amplification,hallaji2017weak}, which is one of the most important applications of weak measurement, aims to estimate the interaction strength of a given weak von Neumann interaction between two systems.
Because the measurement systems for weak-value amplification inevitably involve probe systems, our measurement method appears not applicable to weak-value amplification directly.
In this section, we describe how to apply our measurement method to weak-value amplification.

As shown in Fig.~\ref{fig:4-1}(a), in weak-value amplification, a weak von Neumann interaction between a system to be measured (system 1) and a probe system (system 2) is given, and the unitary transformation caused by this interaction is expressed as $\hat{U}_{12}(\theta)=\exp(\ii\theta\hat{A}_1\otimes\hat{A}_2)$, where $\hat{A}_1$ and $\hat{A}_2$ are Hermite operators on the systems 1 and 2, respectively, and $\theta$ is a small real parameter to be estimated.
To apply our measurement method, we consider the two systems as a composite single system as shown in Fig.~\ref{fig:4-1}(b).
When the total system is pre- and post-selected in $\ket{\varPsi\sub{i}}_{12}=\ket{\ii}_1\ket{\phi\sub{i}}_2$ and $\ket{\varPsi\sub{f}}_{12}=\ket{\ff}_1\ket{\phi\sub{f}}_2$, respectively, the ratio of its post-selection probability $P(\varPsi\sub{f}|\varPsi\sub{i},\theta)$ for $\theta\neq 0$ to that for $\theta=0$ is represented as
\begin{align}
\frac{P(\varPsi\sub{f}|\varPsi\sub{i},\theta)}{P(\varPsi\sub{f}|\varPsi\sub{i},0)}
&=
\frac{|_{12}\bracketii{\varPsi\sub{f}}{\hat{U}_{12}(\theta)}{\varPsi\sub{i}}_{12}|^2}
{|_{12}\bracketi{\varPsi\sub{f}}{\varPsi\sub{i}}_{12}|^2}
\no\\
&=1+2\mathrm{Re}(\ii\bracket{\hat{A}_1}\sub{w}\bracket{\hat{A}_2}\sub{w})\theta+O(\theta^2),
\end{align}
where $\bracket{\hat{A}_1}\sub{w}:={}_1\bracketii{\ff}{\hat{A}_1}{\ii}_1/_1\bracketi{\ff}{\ii}_1$ and $\bracket{\hat{A}_2}\sub{w}:={}_2\bracketii{\phi\sub{f}}{\hat{A}_2}{\phi\sub{i}}_2/_2\bracketi{\phi\sub{f}}{\phi\sub{i}}_2$.
When the pre- and post-selected states are prepared so that $\ii\bracket{\hat{A}_1}\sub{w}\bracket{\hat{A}_2}\sub{w}$ becomes a huge real value, the difference $P(\theta)/P(0)-1$ can be detected experimentally even if $\theta$ is very small, and when $\bracket{\hat{A}_1}\sub{w}$ and $\bracket{\hat{A}_2}\sub{w}$ are known, $\theta$ can be estimated from the difference $P(\theta)/P(0)-1$.
This is the same strategy as weak-value amplification.

In the conventional weak measurement for weak-value amplification, the amount of the probe shift needs to be measured. 
Whereas our measurement method also uses a probe system for weak-value amplification, the post-selected state of the probe system is fixed and the change in the post-selection probability is measured.
Therefore, the implementation of our measurement method is obviously easier than the conventional weak measurement.

\subsection{Accuracy and precision in the proposed measurement method}\label{sec:4-2}

Here, we evaluate the performance of our measurement method as a weak-value estimation method in terms of the accuracy and precision.
First, the accuracy is defined as the difference of the estimator for infinite number of trials from the true value.
In our measurement method, Eq.~(\ref{eq:10}) indicates that there is a difference $O(\theta^2)$ between $\bracketii{\ff}{\hat{N}(\theta)}{\ii}/\bracketi{\ff}{\ii}-1$ and $\bracket{\hat{C}}\sub{w}\theta$, and the estimator of the weak value for infinite number of trials, $(\bracketii{\ff}{\hat{N}(\theta)}{\ii}/\bracketi{\ff}{\ii}-1)/\theta$, contains an error of $O(\theta)$.
Nevertheless, the first order term of $\theta$ in this error can be canceled by measuring not only $\bracketii{\ff}{\hat{N}(\theta)}{\ii}/\bracketi{\ff}{\ii}-1$ but also $\bracketii{\ff}{\hat{N}(-\theta)}{\ii}/\bracketi{\ff}{\ii}-1$ and taking the difference between the two as
\begin{align}
\frac{\bracketii{\ff}{\hat{N}(\theta)}{\ii}}{\bracketi{\ff}{\ii}}
-\frac{\bracketii{\ff}{\hat{N}(-\theta)}{\ii}}{\bracketi{\ff}{\ii}}
=2\bracket{\hat{C}}\sub{w}\theta+O(\theta^3).
\end{align}
If the strength of the transformation $\theta$ is known, $\bracket{\hat{C}}\sub{w}$ can be estimated with an error of $O(\theta^2)$ for infinite number of trials.
This technique can be used for estimating the real and imaginary part of weak values in Eqs.~(\ref{eq:25}) and (\ref{eq:24}), respectively, to cancel the first order term of $\theta$ in their errors.
In weak measurement using a Gaussian probe, the estimation error of a weak value for infinite number of trials can also be reduced to $O(\theta^2)$, where $\theta$ is the interaction strength between the measured and probe systems (see Appendix \ref{sec:appendix2} for details). 

Next, the precision is defined as the uncertainty (mean square error deviation) of the estimator for finite number of trials.
The lower bound of this estimation uncertainty for $n$ trials, $\Delta\bracket{\hat{C}}\sub{w}$, is given by quantum Cram\'{e}r--Rao inequality \cite{PhysRevLett.72.3439} as follows:
\begin{align}
 \Delta\bracket{\hat{C}}\sub{w}\geq \frac{1}{\sqrt{nF}}
\geq \frac{1}{\sqrt{nF\sub{Q}}}, \label{eq:14}
\end{align}
where $F$ and $F\sub{Q}$ are classical and quantum Fisher information, respectively.
When $\theta$ is known and $\mathrm{Re}\bracket{\hat{C}}\sub{w}$ is to be estimated, the classical Fisher information of the probability distribution $\{P(\ff|\ii,\theta)=|\bracketii{\ff}{\hat{N}(\theta)}{\ii}|^2,1-P(\ff|\ii,\theta)\}$ is given as
\begin{align}
 F=\frac{4|\bracketi{\ff}{\ii}|^2}{1-|\bracketi{\ff}{\ii}|^2}\theta^2+O(\theta^3).
\end{align}
This value is larger than that for weak measurement using a Gaussian probe (see Appendix~\ref{sec:appendix2} for details) and becomes particularly large when the denominator $1-|\bracketi{\ff}{\ii}|^2$ is close to zero. 
In contrast, when $\theta$ is to be estimated, the quantum Fisher information of the state $\hat{N}(\theta)\ket{\ii}$ is given as
\begin{align}
 F\sub{Q}=4(\bracketii{\ii}{\hat{C}^\dag\hat{C}}{\ii}-|\bracketii{\ii}{\hat{C}}{\ii}|^2)+O(\theta).
\end{align}
If $\hat{C}$ is anti-Hermite (i.e., $\hat{N}(\theta)$ is unitary), the right equality in Eq.~(\ref{eq:14}) holds when the post-selected state is chosen as
\begin{align}
\ket{\ff}
=\frac{1}{\sqrt{2}}
\left[
\ket{\ii}+
\frac{(\hat{1}-\ket{\ii}\bra{\ii})\hat{C}\ket{\ii}}
{
\|(\hat{1}-\ket{\ii}\bra{\ii})\hat{C}\ket{\ii}\|
}
\right].
\end{align}
We note that the weak value $\bracket{\hat{C}}\sub{w}$ is not huge for this $\ket{\ff}$, which indicates that the strategy of weak-value amplification is not optimal for such an ideal situation, similar to the results in the previous studies \cite{PhysRevA.88.042116,PhysRevLett.112.040406,PhysRevLett.114.210801}.
Nevertheless, there is a possibility that weak-value amplification becomes an optimal strategy under the presence of a specific technical noise \cite{PhysRevLett.118.070802,PhysRevX.4.011031,PhysRevA.92.032127}.

\subsection{Other weak-value measurement methods}\label{sec:4-3}

While our measurement method is an alternative to weak measurement to obtain weak values, some other weak-value measurement methods have been proposed so far \cite{vallone2016strong,denkmayr2017experimental,kedem2010modular,hofmann2014sequential,ogawa2018framework,calderaro2018direct,JOHANSEN2007374,cohen2018determination}.
Here, we mention these weak-value measurement methods other than weak measurement and compare the advantages of each method and our measurement method.

Some of the methods previously reported \cite{vallone2016strong,denkmayr2017experimental,kedem2010modular,hofmann2014sequential,ogawa2018framework,calderaro2018direct} employ indirect (von Neumann) measurement via strong system--probe interactions.
The others \cite{JOHANSEN2007374,cohen2018determination} are direct measurement methods, in which weak values are obtained from a combination of several projective (strong) measurements of pre-selected systems.
These methods have an advantage over weak measurement in terms of efficiency because of the strong interactions or measurements, while the pre- and post-selected systems are strongly disturbed.
Therefore, these methods can be applied to, for example, direct measurements of wavefunctions and pseudo-probability distributions of the system's pre-selected state \cite{lundeen2011direct,lundeen2012procedure,salvail2013full,kobayashi2014stereographical,malik2014direct,mirhosseini2014compressive,shi2015scan,thekkadath2016direct}.

In contrast, the characteristic of hardly disturbing measured pre- and post-selected systems, which weak measurement has, is essential for the studies that have investigated physical quantities of quantum systems after post-selection by weak measurement \cite{aharonov1991complete,resch2004experimental,PhysRevLett.102.020404,yokota2009direct,kocsis2011observing,goggin2011violation,denkmayr2014observation,mahler2016experimental} and weak-value amplification \cite{hosten2008observation,dixon2009ultrasensitive,magana2014amplification,hallaji2017weak}.
Our weak-value measurement method also maintains the characteristic of hardly disturbing the pre- and post-selected systems because the weak values are obtained at the limit of very small transformation.
Therefore, our measurement method is widely applicable to such applications instead of weak measurement.

\section{Conclusion}\label{sec:5}



In this study, we proposed an operational formulation of weak values as the response of the pre- and post-selected system, without using probe systems. 
In this formulation, when a small quantum transformation is given between the pre- and post-selections, the weak value of the generator of this small transformation appears in the change of the post-selection probability amplitude.
This formulation is a generalization of the results that have been reported before and therefore covers various cases, such as when the small transformation is other than unitary or attenuation transformations, or when the pre- and post-selection is in a mixed state.
We applied this formulation to examples of the quantum box problem and the huge weak value of spin to provide a direct interpretation of the strange weak values as a natural extension of the conventional physical quantities such as probability and spin angular momentum.
We also explained that this measurement method can be applied to simplify various weak-value measurement experiments.
Thus, the proposed operational formulation of weak values, freed from the concept of probe shift, is expected to play an important role in both fundamental and practical investigation on weak values and pre- and post-selection quantum systems.

\begin{acknowledgments}
We thank Y.~Shikano for the useful comments and discussion. 
This research was supported by JSPS KAKENHI Grant Number 16K17524 and 19K14606, the Matsuo Foundation, and the Research Foundation for Opto-Science and Technology.
\end{acknowledgments}

\appendix

\section{How to realize amplification/attenuation transformation}\label{sec:appendix1}

In this appendix, we describe how to realize the small amplification/attenuation transformation by unitary processes.
While the time evolution of closed systems must be unitary in quantum mechanics, the small amplification/attenuation transformation in a target space can be realized effectively by embedding that transformation into a unitary transformation on a dilated space, as shown below.
Let us consider the target $d$-dimensional Hilbert space $\mathcal{H}$ and a dilated $(d+d')$-dimensional Hilbert space $\mathcal{H}\oplus\mathcal{H}'$, where $\mathcal{H}'$ is an additional $d'$-dimensional Hilbert space satisfying $\mathcal{H}'\perp\mathcal{H}$. 
The initial state $k(\ket{\ii}+\ket{\chi\sub{i}})$, where $\ket{\ii}\in\mathcal{H}$, $\ket{\chi\sub{i}}\in\mathcal{H}'$, and $k=1/\sqrt{\bracketi{\ii}{\ii}+\bracketi{\chi\sub{i}}{\chi\sub{i}}}$, undergoes unitary evolution $\hat{U}(\theta)=\hat{1}+\ii\theta\hat{A}+O(\theta^2)$, where $\hat{A}$ is a Hermite operator on $\mathcal{H}\oplus\mathcal{H}'$ and then is projected onto $\mathcal{H}$ by $\hat{P}:=\hat{1}_\mathcal{H}\oplus 0_{\mathcal{H}'}$.
$\hat{A}$ is represented as $\hat{A}=
\begin{bmatrix}
 \hat{A}_{11}&\hat{A}_{12}\\
\hat{A}_{21}&\hat{A}_{22}\\
\end{bmatrix}$,
where $\hat{A}_{11}:\mathcal{H}\rightarrow\mathcal{H}$ and $\hat{A}_{22}:\mathcal{H'}\rightarrow\mathcal{H'}$ are Hermite operators and $\hat{A}_{12}:\mathcal{H'}\rightarrow\mathcal{H}$ and $\hat{A}_{21}:\mathcal{H}\rightarrow\mathcal{H'}$ satisfy $\hat{A}_{12}^\dag=\hat{A}_{21}$ due to the Hermitian condition $\hat{A}^\dag=\hat{A}$.
The resulting state after the transformation is given as
\begin{align}
&\hat{P}\hat{U}(\theta)k(\ket{\ii}+\ket{\chi\sub{i}})\no\\
&=(\hat{1}+\ii\theta\hat{A}_{11})k\ket{\ii}+\ii\theta\hat{A}_{12}k\ket{\chi\sub{i}}+O(\theta^2).\label{eq:8}
\end{align}
The first term of the right-hand side, $(\hat{1}+\ii\theta\hat{A}_{11})k\ket{\ii}$, represents the unitary evolution in $\mathcal{H}$, and the second term, $\ii\theta\hat{A}_{12}k\ket{\chi\sub{i}}\in\mathcal{H}$, represents the inflow of the amplitude from $\mathcal{H}'$ to $\mathcal{H}$.
Therefore, if  $\hat{A}_{11}=0$ and $\ii\theta\hat{A}_{12}k\ket{\chi\sub{i}}$ is represented as $\hat{B}k\ket{\ii}$, the amplification/attenuation transformation $\hat{1}+\theta\hat{B}+O(\theta^2)$ for the state $k\ket{\ii}$ is effectively realized.


Another way to realize an amplification/attenuation transformation is using an interaction with an ancilla system.
Let us consider a weak von Neumann interaction between a system to be measured (system 1) and a probe system (system 2) expressed as $\exp(\ii\theta\hat{A}_1\otimes\hat{A}_2)$, where $\hat{A}_1$ and $\hat{A}_2$ are Hermite operators on the systems 1 and 2, respectively.
When the system 2 is pre- and post-selected in $\ket{\phi\sub{i}}_2$ and $\ket{\phi\sub{f}}_2$, respectively, the system 1 undergoes the following effective small transformation: 
\begin{align}
{}_2\bra{\phi\sub{f}}&{\exp(\ii\theta\hat{A}_1\otimes\hat{A}_2)}\ket{\phi\sub{i}}_2\no\\
&={}_2\bracketi{\phi\sub{f}}{\phi\sub{i}}_2
\big[
1+\theta(\ii \bracket{\hat{A}_2}\sub{w}\hat{A}_1)+O(\theta^2)
\big],
\end{align}
where $\bracket{\hat{A}_2}\sub{w}:={}_2\bracketii{\phi\sub{f}}{\hat{A}_2}{\phi\sub{i}}_2/{}_2\bracketi{\phi\sub{f}}{\phi\sub{i}}_2$.
The kind of the effective small transformation depends on $\bracket{\hat{A}_2}\sub{w}$.  
If $\bracket{\hat{A}_2}\sub{w}$ is real, the effective small transformation becomes a unitary transformation whose generator is an anti-Hermite operator $\ii\bracket{\hat{A}_2}\sub{w}\hat{A}_1$.
In contrast, if $\bracket{\hat{A}_2}\sub{w}$ is purely imaginary, the effective small transformation becomes an amplification/attenuation transformation whose generator is a Hermite operator $\ii\bracket{\hat{A}_2}\sub{w}\hat{A}_1$.
We note that the case where the norm of the state vector of system 1 exceeds one because of the effective amplification transformation is understood as follows:
the post-selection probability for system 2 becomes larger than $|_2\bracketi{\phi\sub{f}}{\phi\sub{i}}_2|^2$; therefore, the number of trials of system 1 that remains after the post-selection of system 2 is increased.


\section{Accuracy and precision in weak measurement using a Gaussian probe}\label{sec:appendix2}

In this appendix, we provide the accuracy and precision in weak measurement using a Gaussian probe.
Before that, we review the weak measurement using a Gaussian probe.
The initial states of the measured and probe systems are $\ket{\ii}$ and $\ket{\phi}$, respectively.
We assume that $\ket{\phi}$ can be expanded in the position basis $\{\ket{x}\}$ as
\begin{align}
 \ket{\phi}&=\int_{-\infty}^{\infty}\dd x\phi(x)\ket{x},\quad\phi(x)=\frac{1}{\pi^{1/4}\sqrt{\sigma}}\exp\frac{-x^2}{2\sigma^2}.
\end{align}
The initial state $\ket{\ii}\ket{\phi}$ is evolved through the weak system--probe interaction $\exp(-\ii\theta\hat{A}\otimes\hat{p})$, where $\hat{A}$ is the measured observable, $\hat{p}$ is the momentum operator of the probe system, and $\theta$ is a small coupling constant satisfying $\theta\|\hat{A}\|\ll \sigma$ ($\|\hat{A}\|$ is the maximum eigenvalue of $\hat{A}$). 
After the system is post-selected into $\ket{\ff}$, the unnormalized state of the probe system $\ket{\tilde{\phi}\sub{f}}$ is represented as 
\begin{align}
\ket{\tilde{\phi}\sub{f}}
&=\bracketii{\ff}{\exp(-\ii\theta\hat{A}\otimes\hat{p})}{\ii}\ket{\phi}\no\\
&=\bracketi{\ff}{\ii}\left(\hat{1}-\ii\theta\bracket{\hat{A}}\sub{w}\hat{p}-\frac{\theta^2}{2}\bracket{\hat{A}^2}\sub{w}\hat{p}^2\right)\ket{\phi}+O(\theta^3).
\end{align}
The projection measurement of the position $\{\ket{x}\bra{x}\}$ for $\ket{\tilde{\phi}\sub{f}}$ gives the following probability density distribution $P(x|\mathrm{Re}\bracket{\hat{A}}\sub{w})$:
\begin{widetext}
\begin{align}
P(x|\mathrm{Re}\bracket{\hat{A}}\sub{w})
=\frac{\bracketi{\tilde{\phi}\sub{f}}{x}\bracketi{x}{\tilde{\phi}\sub{f}}}{\bracketi{\tilde{\phi}\sub{f}}{\tilde{\phi}\sub{f}}}
=|\phi(x)|^2\left[
1+\frac{2\mathrm{Re}\bracket{\hat{A}}\sub{w}x}{\sigma^2}\theta
+\frac{|\bracket{\hat{A}}\sub{w}|^2+\mathrm{Re}\bracket{\hat{A}^2}\sub{w}}{2\sigma^2}
\left(\frac{x^2}{\sigma^2}-\frac{1}{2}\right)\theta^2
\right]+O(\theta^3).
\end{align}
\end{widetext}

In weak measurement, $\mathrm{Re}\bracket{\hat{A}}\sub{w}$ is estimated as the averaged value of the outcomes of the position measurement to the probe system.
For infinite number of trials, the estimator of $\mathrm{Re}\bracket{\hat{A}}\sub{w}$ is given as the expectation value of the probability density distribution $P(x|\mathrm{Re}\bracket{\hat{A}}\sub{w})$, which is represented as
\begin{align}
\int_{-\infty}^\infty\dd x P(x|\mathrm{Re}\bracket{\hat{A}}\sub{w})
=\theta\mathrm{Re}\bracket{\hat{A}}\sub{w}+O(\theta^3).
\end{align}
Therefore, $\mathrm{Re}\bracket{\hat{A}}\sub{w}$ can be estimated with an error of $O(\theta^2)$ for infinite number of trials.

The classical Fisher information of the probability density distribution $P(x|\mathrm{Re}\bracket{\hat{A}}\sub{w})$ is given as
\begin{align}
 J&=\int_{-\infty}^\infty\dd x
\frac{1}{P(x|\mathrm{Re}\bracket{\hat{A}}\sub{w})}
\left[
\frac{\dd P(x|\mathrm{Re}\bracket{\hat{A}}\sub{w})}
{\dd(\mathrm{Re}\bracket{\hat{A}}\sub{w})}
\right]^2\no\\
&=|\bracketi{\ff}{\ii}|^2\theta'^2+O(\theta'^3),
\end{align}
where $\theta':=\theta/\sigma$ is the parameter representing the substantial strength of the interaction.

\end{spacing}

\nocite{*}
\bibliography{ref}

\end{document}